\documentstyle[12pt]{article}
    
\renewcommand{\theequation}{\arabic{section}.\arabic{equation}}

\renewcommand{\thesection}{\arabic{section})}

 \newcommand{\be}{\begin{equation}}
 \newcommand{\ee}{\end{equation}}
 \newcommand{\ba}{\begin{eqnarray}}
 \newcommand{\ea}{\end{eqnarray}}
 
 \newcommand{\del}{\partial}

\def\B{\tilde B}

\def\C{\tilde C}

\newcommand{\expo}{\exp \lef \{ - \int d^3z }

\newcommand{\lef}{\left}

\newcommand{\ri}{\right}

\newcommand{\cl}{{\cal L}}

\newcommand{\fr}{\frac}

\newcommand{\emab}{\epsilon^{\mu\alpha\beta}}

\begin{document}

\begin{titlepage}

\topmargin -15mm

\vskip 10mm

\centerline{ \LARGE\bf Charge and Magnetic Flux Correlations}
\vskip 2mm
\centerline{ \LARGE\bf in Chern-Simons Theory with Fermions}

    \vskip 2.0cm

    \centerline{\sc E.C.Marino }

     \vskip 0.6cm
     
    \centerline{\it Instituto de F\'\i sica }
    \centerline{\it Universidade Federal do Rio de Janeiro }
    \centerline{\it Cx.P. 68528}
    \centerline{\it Rio de Janeiro RJ 21945-970}
    \centerline{\it Brasil}

\vskip 1.0cm

\begin{abstract} 
 
Charge and magnetic flux bearing operators are introduced in Chern-Simons
theory both in its pure form and when it is coupled to fermions. The 
magnetic flux creation operator turns out to be the Wilson line. The
euclidean correlation functions of these operators are shown to be local
and are evaluated exactly in the pure case and through an expansion in the 
inverse fermion mass whenever these are present. Physical states only
occur in the presence of fermions and consist of composite charge-magnetic 
flux carrying states which are in general anyonic. The large distance
behavior of the correlation functions indicates the condensation of
charge and magnetic flux.

\end{abstract}

\vskip 3cm
$^*$ Work supported in part by CNPq-Brazilian National Research Council.
     E-Mail address: marino@if.ufrj.br

\end{titlepage}

\hoffset= -10mm

\leftmargin 23mm

\topmargin -8mm
\hsize 153mm
 
\baselineskip 7mm
\setcounter{page}{2}

\section{Introduction}

Chern-Simons theory presents several peculiar properties which made it
the object of intense investigation in the few past years. Besides the
interest in the purely formal aspects of it there is also a vast 
potential of phenomenological applications in the physics of 
condensed matter, more specifically in the quantum Hall effect and high
temperature superconductivity. 

In this work we consider Chern-Simons theory, both in its pure form 
and when it is coupled to fermions
and investigate the properties of some coherent-like states of the 
gauge field which
carry either magnetic flux, charge or both. The idea is to construct
the creation operators for such states by following a general procedure
which was applied before \cite{nv,ap,evora} 
(see also \cite{rede,cont} for related 
works) in theories containing Maxwell terms  
in the lagrangian. As it turns out, the magnetic flux (vortex) creation
operator is the Wilson line and its correlation functions are proved
to be local (line independent) in Chern-Simons theories. 
In the absence of fermions, the
quantum vortex states are unphysical, 
because they do not satisfy the condition
imposed by the field equation on physical states. When fermions are 
coupled to the Chern-Simons field, we show that the only physical states 
are the ones created by composite charge-magnetic-flux 
carrying operators which are in general anyonic, except for a 
specific value of the $\theta$ parameter of the Chern-Simons theory. We
evaluate the correlation functions of them through an expansion in
the inverse fermion mass. The large distance behavior of these indicate
a total condensation of charge and magnetic flux.
						 
In Section 2  we consider pure Chern-Simons theory and introduce  
creation operators for vortex-like  
excitations examining the condition for locality of their correlation 
functions. The relevant commutators of these operators
as well as their correlation functions are evaluated. 
We also introduce the $\sigma$-
operators which, in the presence of fermions, will 
become the creation operators of charged states.

In Section 3, we couple fermions to the theory and 
determine the corresponding form of the operators
carrying charge and magnetic flux under the condition of locality of their
correlation functions. We also evaluate the basic commutators and 
the explicit form of the correlation functions. 
	    
Perspectives of application  for the work developed here are
presented in Section 4. Four Appendixes are included to demonstrate 
useful results.

\section{Pure Chern-Simons Theory}

\subsection{Flux Carrying Operators}
\setcounter{equation}{0}

Let us consider in this section pure Chern-Simons theory, whose lagrangian
is given by
\be
\cl = \frac{\theta}{4} \epsilon^{\mu\alpha\beta}A_\mu F_{\alpha\beta}
\label{cs}
\ee
where $\theta$ is an arbitrary real number. We want to obtain an 
operator which creates states bearing magnetic flux and whose correlation
functions are local.
In a previous work \cite{nv},
we showed that in Maxwell-type theories such an operator could be 
obtained by requiring that its correlation functions were given by the
vacuum functional in the presence of the external field
\be
\B^{\mu\nu}(z;x) = b \int_{x,L}^\infty d\xi_\lambda 
\epsilon^{\lambda\mu\nu}
\delta^3(z-\xi)
\label{b}
\ee
which would be coupled through the shift
$F^{\mu\nu} \rightarrow F^{\mu\nu} + \B^{\mu\nu}$. The resulting 
correlation functions were shown to be independent of the curve $L$
appearing in the definition of $\B_{\mu\nu}$ 
in (\ref{b}), being therefore local.
In the same way, we are going to introduce a magnetic flux carrying 
operator in Chern-Simons theory by expressing its 
euclidean correlation function as
\be
<\mu(x) \mu^\dagger(y)> = Z^{-1} \int DA_\mu \expo \lef[ -\fr{i\theta}{4}
\emab A_\mu (F_{\alpha\beta} + 2 \B_{\alpha\beta}(x,y) ) \ri]\ri\}
\label{cf1}
\ee
where $\B(x,y)= \B(z,x)-\B(z,y)$ with $\B(z,x)$
given by (\ref{b}). Notice the factor 2 in the 
above expression. As we will see, this will compensate for the fact that
the Chern-Simons lagrangian is not completely expressed in terms of 
$F_{\mu\nu}$ as in the case of Maxwell.

From the above expression we can extract the 
form of the flux carrying operator,
namely
$$
\mu(x) = \exp \lef\{\fr{i\theta}{2} \int d^3 z 
\emab A_\mu \B_{\alpha \beta}\ri\}
$$
\be
\mu(x) = \exp \lef\{-ib\theta \int_{x,L}^\infty d\xi^\mu A_\mu  \ri\}
\label{mu1}
\ee
where in the last step we already went back to Minkowski space.
We see that $\mu$ is the Wilson line 
operator. We will show that in Chern-Simons theory it creates sates
carrying magnetic flux and possesses local ($L$-independent) correlation
functions.
 
We can show the path independence of (\ref{cf1}) by making the change of
variable
$$
A_\mu \rightarrow A_\mu - \Omega_\mu
$$
\be
\Omega_\mu = b  \int_{S(L,L')} d^2\xi_\mu \delta^3( z-\xi) 
\label{ome}
\ee
where $L'$ is an arbitrary path going from $x$ (or $y$) 
to infinity, $S(L,L')$
is the surface bounded by $L$ and $L'$
and $d^2\xi^\mu$ its surface element. Under the above change of variable, 
the exponent in (\ref{cf1}) changes as (see Appendix A)
\be
S_{CS}[\B_{\mu\nu}(L)] \rightarrow S_{CS}[\B_{\mu\nu}(L')]
\label{ds}
\ee
and threfore the path independence of $\mu$-correlation functions is 
established.

Let us evaluate now the equal time commutator of $\mu$ 
with the magnetic flux 
operator 
$$
\Phi = \int d^2x \epsilon^{ij} \del_i A_j
$$
Using (\ref{mu1}), we have
$$
[\mu(x),\Phi] = \mu(x)(-ib\theta)\int d^2y \int_{x,L}^\infty d\xi^k
\epsilon^{ij} \del^{(y)}_i [A_k(\xi), A_j(y)]
$$
\be
\ \ \ \ = \mu(x) b \int d^2y \int_{x,L}^\infty d\xi^i  \del^{(y)}_i
\delta^2(\xi -y) = b \mu(x)
\label{com1}
\ee
where we used the equal-time commutator $[A_i(x),A_j(y)]=\fr{i}{\theta}
\epsilon^{ij} \delta^2(\vec x -\vec y)$.
This shows that indeed the $\mu$-operator introduced above carries $b$
units of magnetic flux. It is therefore a ``vortex'' creation operator.

\subsection{$\sigma$-Operators}

Let us intruduce now an operator called $\sigma$ which plays an 
interesting role, in connection with the magnetic flux carrying
operator $\mu$. In Maxwell theory \cite{oop} $\sigma$ would 
create charged states, being  
the dual of $\mu$, in the sense of order-disorder duality. Also
when coupling charged fermions to the Chern-Simons field (see next
Section) we will see that $\sigma$ becomes a charge carrying operator.
Most important of all, however, is the fact that the physical states of the
theory will be created by the composite operator $\sigma\mu$.
		   
Again, following the procedure
taken for $\mu$, we will introduce local $\sigma$ correlation functions
by coupling to $F_{\mu\nu}$ an external field which was used 
in Maxwell type theories in order to describe the analogous charged 
operators \cite{oop}, namely
$$
\C^{\alpha\beta}(z;x) = \del^\alpha \C^\beta(z;x) 
- \del^\beta \C^\alpha(z;x)
$$
with
\be
\C^\alpha(z;x) = i a \int_{x,L}^\infty d\xi^\alpha \delta^3 (z- \xi)
\label{c}
\ee
We introduce the euclidean $\sigma$-correlation functions through
\be
<\sigma(x) \sigma^\dagger(y)> = Z^{-1} \int DA_\mu \expo 
\lef[ -\fr{i\theta}{4}
\emab A_\mu (F_{\alpha\beta} + 2 \C_{\alpha\beta}(x,y) ) \ri]\ri\}
\label{cf2}
\ee
where $\C(x,y)= \C(z,x)-\C(z,y)$ with $\C(z,x)$ given by (\ref{c}). 

Now, from (\ref{cf2}) we can derive the form of the $\sigma$-operator
$$
\sigma(x) = \exp \lef \{ ia\theta \int d^3z \emab A_\mu \del_\alpha
\C_\beta \ri \}
$$
\be
\sigma(x) = \exp \lef \{ ia\theta \int_{x,L}^\infty d\xi^i
\epsilon^{ij} E_j \ri \} 
\label{si}
\ee
where we went back to Minkowski space in the last step and chose a 
spatial path. $E^i$ is the electric field.
From the second expression for $\sigma$ in (\ref{si}) we see that it
commutes with the magnetic flux operator $\Phi$.

As for the case of $\mu$, let us show the path
independence of (\ref{cf2}). We do this by making in (\ref{cf2})
the change of variable
$$
A_\mu \rightarrow A_\mu - \tilde\Omega_\mu
$$
\be
\tilde\Omega_\mu = a \oint_{L'-L} d\xi_\mu \delta^3( z-\xi) 
\label{omet}
\ee
where $L'$ is an arbitrary path going from $x$ (or $y$) 
to infinity.
Under the above change of variable, 
the exponent in (\ref{cf2}) changes as (see Appendix B)
\be
S_{CS}[\C_{\mu\nu}(L)] \rightarrow S_{CS}[\C_{\mu\nu}(L')]
\label{ds5}
\ee
and therefore the correlation function (\ref{cf2}) is path independent.

\subsection{Correlation Functions}

Let us evaluate in this subsection the correlation functions involving the 
operators $\sigma$ and $\mu$
introduced previously. We start with the mixed correlation function which
according to (\ref{cf1}) and (\ref{cf2}) is given by
$$
<\sigma(x_1) \mu(x_2) \sigma^\dagger(y_1)\mu^\dagger(y_2)> = 
Z^{-1} \int DA_\mu \expo 
\lef[ -\fr{i\theta}{4}
\emab A_\mu (F_{\alpha\beta} +
\ri .\ri .
$$
\be
\lef . \lef .
 2\B_{\alpha\beta}(x_2,y_2) 
+ 2 \C_{\alpha\beta}(x_1,y_1) ) \ri]\ri\}
\label{cf3}
\ee
This quadractic functional integral can be evaluated through the use of the
euclidean propagator of the Chern-Simons field,
\be
D^{\mu\nu}(x) = \fr{i}{\theta} \epsilon^{\mu\alpha\nu} \del_\alpha
\fr{1}{4\pi |x|} + gauge \  terms
\label{pro}
\ee
giving the result
\be
<\sigma\mu \sigma^\dagger\mu^\dagger> = \exp \lef\{ \fr{i}{2\theta}
\int d^3z d^3z' [B_\mu + D_\mu](z) [B_\nu + D_\nu](z') 
\epsilon^{\mu\alpha\nu} \del_\alpha \lef [ \fr{1}{4\pi|z-z'|}\ri ]\ri\}
\label{cf4}
\ee
where
$$
B_\mu = \fr{-i\theta}{2} \emab [\B_{\alpha\beta}(x_2)-\B_{\alpha\beta}(y_2)]
= -i b\theta \int_{x_2,L}^{y_2} d\xi_\mu \delta^3(z-\xi)
$$
\be
D_\mu = -i\theta \emab \del_\beta [\C_{\alpha}(x_1)-\C_{\alpha}(y_1)]
= a \theta \int_{x_1,L}^{y_1} d\xi_\lambda  
\epsilon^{\lambda\alpha\mu} \del_\alpha  \delta^3(z-\xi)
\label{bc}
\ee
The exponent in (\ref{cf4}) will have three terms: $BB$, $DD$ and $BD$. Let
us first consider the 
$DD$ term. Inserting $D_\mu$, integrating over $z$ and $z'$ and using
a trivial identity for the $\epsilon$'s we get
\be
DD = -i a^2 \theta \int_{x_1,L}^{y_1} d\xi_\lambda
\int_{x_1,L}^{y_1} d\eta_\mu
\epsilon^{\lambda\alpha\mu} \del_\alpha  \delta^3(\xi-\eta) = 0
\label{dd}
\ee
This is zero because $d\xi_\lambda \parallel d\eta_\mu$ for $\xi =\eta$.
Let us turn now to the $BD$ term. Inserting (\ref{bc}) in (\ref{cf4}),
and following the same steps which led to (\ref{dd}) we get 
\be
BD = a b \theta \int_{x_2,L}^{y_2} d\xi_\mu
\int_{x_1,L'}^{y_1} d\eta_\nu
[ \delta^{\mu\nu} (-\Box) - \del^\mu_{(\xi)} \del^\nu_{(\eta)}
\lef [ \fr{1}{4\pi |\xi -\eta|} \ri ] 
\label{bd}
\ee
The expression between brackets is the Green function of $-\Box$. Hence,
the first term above is proportional to $\delta^3(\xi-\eta)$ and 
vanishes because we can always choose the curves $L$ and $L'$ as 
nonintercepting. Evaluation of the second term is trivial and gives
\be
BD =\theta \fr{ab}{4\pi} \lef [\fr{1}{|y_1-x_2|} + \fr{1}{|y_2 -x_1|}
-\fr{1}{|y_2-y_1|} - \fr{1}{|x_2 - x_1|} \ri ] 
\label{bd1}
\ee
Now for the $BB$ term. Using (\ref{b}) and (\ref{pro}) we obtain
\be
BB = - \fr{ib^2\theta}{2} \int_{x_2,L}^{y_2} d\xi_\mu
\int_{x_2,L}^{y_2} d\eta_\nu \epsilon^{\mu\alpha\nu} \del_\alpha^{(\xi)}
\lef [ \fr{1}{4\pi |\xi -\eta|} \ri ]
\label{bb}
\ee
In Appendix C we show that this expression gives
\be
BB = -i\fr{b^2 \theta}{4\pi} [ arg(\vec x_2-\vec y_2) 
+ arg(\vec y_2-\vec x_2)
- 2 arg(\vec \epsilon)  ]
\label{bb1}
\ee
where $\epsilon$ is a regulator which we must set to zero.
			  
Collecting all terms contributing to (\ref{cf4}), we get
$$
<\sigma(x_1) \mu_R(x_2) \sigma^\dagger(y_1)\mu_R^\dagger(y_2)> = \exp \lef\{ 
\theta \fr{ab}{4\pi} \lef [\fr{1}{|y_1-x_2|} + \fr{1}{|y_2 -x_1|}
-\fr{1}{|y_2-y_1|} - \fr{1}{|x_2 - x_1|} \ri ] \ri .
$$
\be
\lef .
 -i\fr{b^2 \theta}{4\pi} [ arg(\vec x_2-\vec y_2) + arg(\vec y_2-\vec x_2)]
\ri \}
\label{cf5}
\ee
where we defined the renormalized fields
\be
\mu_R = \exp \lef\{ i\fr{b^2 \theta}{4\pi} arg(\vec \epsilon)\ri\} \mu
\label{mur}
\ee

From this point it is very easy to obtain the $<\mu \mu^\dagger>$
and $<\sigma \sigma^\dagger>$ correlation functions. They are going to
be the exponentials of the $BB$ and $DD$ terms, respectively. So we get
right away
\be
<\mu_R(x) \mu_R^\dagger(y)> = \exp \lef\{ 
-i\fr{b^2 \theta}{4\pi} [ arg(\vec x-\vec y) + arg(\vec y-\vec x)]\ri\} 
\label{cf6}
\ee
and
\be
<\sigma(x) \sigma^\dagger(y)> = 1 
\label{cf7}
\ee

An important remark is now in order, concerning the states whose
correlation functions we are evaluating.
In pure Chern-Simons theory the field equation
\be
\theta \emab \del_\alpha A_\beta = 0
\label{feq}
\ee
implies that we must impose the ``Gauss' Law'' condition 
$$
\epsilon^{ij} \del_i A_j |{\rm phys} > =0
$$
or
\be   
\Phi |{\rm phys}> =0
\label{cons}
\ee
on the physical states $|{\rm phys}>$. We automatically conclude that the
$|\mu>$ states are not physical by virtue of (\ref{com1}). We could also
consider the states created by the composite operator $\sigma\mu$, whose
correlation functions would be easily obtained from (\ref{cf5}) but
we immediately see that those would also not satisfy (\ref{cons}), 
being therefore unphysical as well. When we couple fermions to the Chern-
Simons theory, on the contrary, we will see in the next Section that
the only physical states will be presicely the ones created by $\sigma\mu$.
There we will study more carefully the correlation functions and properties
of these composite operators.

\section{Chern-Simons Theory with Fermions}
\setcounter{equation}{0}

\subsection{Charge and Magnetic Flux Carrying Operators}
\setcounter{equation}{0}  

Let us consider now the Chern-Simons theory coupled to a Dirac fermion,
\be
\cl = \frac{\theta}{4} \epsilon^{\mu\alpha\beta}A_\mu F_{\alpha\beta}
+ i \bar\psi \not\del \psi - M \bar \psi \psi - e \bar \psi 
\gamma^\mu \psi A_\mu
\label{csf}
\ee
The field equation is 
\be
\theta \emab \del_\alpha A_\beta = e j^\mu
\label{feq1}
\ee
where $j^\mu = \bar \psi \gamma^\mu \psi$.
As a consequence the physical states must satisfy now the condition
$$
\theta\epsilon^{ij} \del_i A_j |{\rm phys} > = e j^0 |{\rm phys} >
$$
or
\be   
\theta\Phi |{\rm phys}> = e Q |{\rm phys} >
\label{cons1}
\ee
where $Q = \int d^2x j^0$ is the matter charge operator.

In order to obtain the local operators $\sigma$ and $\mu$
carrying respectively charge and magnetic flux 
in the presence of fermions, we shall follow the same procedure adopted
previously and introduce the external fields $\B_{\mu\nu}$ and $\C^\alpha$
in such a way that the correlation functions involving $\sigma$ and $\mu$
are local. Starting with the $\sigma$-correlation function, we write
$$
<\sigma(x) \sigma^\dagger(y)> = Z^{-1} \int DA_\mu \expo 
\lef[ -\fr{i\theta}{4}
\emab A_\mu (F_{\alpha\beta} + 2 \C_{\alpha\beta}(x,y) ) 
\ri .\ri .
$$
\be
\lef .\lef .
+ i \bar\psi \not\del \psi - M \bar \psi \psi + e \bar \psi 
\gamma^\mu \psi \lef [ A_\mu + \C_\mu(x,y) \ri ]
\ri]\ri\}
\label{cff2}
\ee
where $\C(x,y)= \C(z,x)-\C(z,y)$ with $\C(z,x)$ given by (\ref{c}). 
This expression is easily seen to be path independent because of
(\ref{omet}) and (\ref{ds5}).

Let us turn now to the $\mu$-operator. For this, we must couple the
external field $\B_{\mu\nu}$ by adding it to $F_{\mu\nu}$. Since the 
action now is not completely expressed in terms of $F_{\mu\nu}$ because
of the interaction term, we are going to use the gauge invariance of the 
fermionic determinant (at least under ``small'' gauge transformations)
to write
$$
<\mu(x) \mu^\dagger(y)> = Z^{-1} \int DA_\mu \expo \lef[ -\fr{i\theta}{4}
\emab A_\mu (F_{\alpha\beta} + 2 \B_{\alpha\beta}(x,y) ) 
\ri . \ri .
$$
\be
\lef .\lef .
+ i \bar\psi \not\del \psi - M \bar \psi \psi + e \bar \psi 
\gamma_\mu \psi \lef [ \fr{\del_\nu 
 [ F^{\mu\nu} + \B^{\mu\nu} ]}{-\Box} \ri ]
\ri]\ri\}
\label{cff1}
\ee
where $\B(x,y)= \B(z,x)-\B(z,y)$ with $\B(z,x)$
given by (\ref{b}). Notice that the action used above differs from 
the one corresponding to (\ref{csf}) by a gauge transformation and are 
therefore equivalent.

The path independence of (\ref{cff1}) follows immediately from (\ref{ome}),
(\ref{ds}) and the fact that under (\ref{ome})
\be
F^{\mu\nu} \rightarrow F^{\mu\nu} + \B^{\mu\nu}[L'] - \B^{\mu\nu}[L]
\label{df}
\ee
From (\ref{cff2}) and (\ref{cff1}), respectively, we can infer the 
form of the $\sigma$ and $\mu$ operators in the presence of fermions, 
namely,
\be
\sigma(x) = \exp \lef \{ ia\theta \int_{x,L}^\infty d\xi^i
\epsilon^{ij} E_j + i e a \theta \int_{x,L}^\infty 
d\xi^\mu j_\mu 
\ri \} 
\label{sif}
\ee
and
\be
\mu(x) = \exp \lef\{-ib\theta \int_{x,L}^\infty d\xi^\mu A_\mu  
-ieb\theta \int_{x,L}^\infty d\xi_\lambda \epsilon^{\lambda\mu\nu}
\del_\nu \fr{j_\mu}{-\Box}
\ri\}
\label{muf1}
\ee
Observe that if one imposed the field equation (\ref{feq1}) on the 
above operators one would obtain the result that the  first and second
terms in the exponents of (\ref{sif}) and (\ref{muf1}) are identical.

Let us investigate now the commutation rules of the $\sigma$ and $\mu$ 
operators obtained above. It is clear that the commutators of these
operators with $\Phi$ remain the same as the ones found in 
Section 2. In order to evaluate the commutator of $\sigma$ with the 
matter charge operator $Q$, let us consider the current algebra relation 
\be
[j^0(\vec x,t), j^i(\vec y, t) ] = i {\cal M} \del^i \delta^2 (\vec x -
\vec y)
\label{jj}
\ee
where ${\cal M}$ is a functional of the spectral density of the theory
and has dimension of mass.
Using this, we easily get
$$
[j^0(\vec y,t), \sigma (\vec x, t) ] = e a \theta
{\cal M} \sigma (\vec x, t)
\delta^2 (\vec x - \vec y) 
$$
or
\be
[ Q, \sigma (\vec x, t) ] = e a \theta {\cal M} \sigma (\vec x, t)
\label{comq}
\ee
The choice $a^{-1}= e \theta {\cal M}$ would imply 
that the operator $\sigma$ carried
one unit of electric charge.

In Appendix D we show that the operator $\mu$ in the presence 
of fermions, which is given by (\ref{muf1}) commutes with the 
charge operator.

\subsection{The Mixed Correlation Function 
$<\sigma \mu \sigma^\dagger \mu^\dagger>$}

Let us evaluate now the correlation functions of the charge and magnetic
flux carrying operators in the presence of fermions. Using expressions
(\ref{cff2}) and (\ref{cff1}) we write
$$
<\sigma(x_1) \mu(x_2) \sigma^\dagger(y_1)\mu^\dagger(y_2)> =
 Z^{-1} \int DA_\mu \expo 
\lef[ -\fr{i\theta}{4}
\emab A_\mu \lef (F_{\alpha\beta} \ri. \ri.\ri.
$$
\be
\lef. \lef.\lef.
+ 2 \B_{\alpha\beta}(x_2,y_2) +
2 \C_{\alpha\beta}(x_1,y_1)\ri) 
+ i \bar\psi \not\del \psi - M \bar \psi \psi + e \bar \psi 
\gamma^\mu \psi \lef [ A_\mu + \C_\mu + 
\fr{\del_\nu \B^{\mu\nu}}{-\Box}
\ri ]
\ri]\ri\}
\label{cff3}
\ee
where we again used the gauge invariance of the fermionic determinant.
Let us now integrate over the fermionic field using an expansion in 
inverse powers of the mass $M$. The leading contribution gives \cite{dr}
$$
<\sigma \mu \sigma^\dagger \mu^\dagger> =
 Z^{-1} \int DA_\mu \expo 
\lef[ -\fr{i\theta}{4}
\emab A_\mu \lef (F_{\alpha\beta} + 2 \B_{\alpha\beta} +
2 \C_{\alpha\beta}\ri) 
\ri.\ri.
$$
\be
\lef.\lef.
-i\fr{e^2}{16\pi} \emab
 \lef [ A_\mu + \C_\mu + 
\fr{\del_\nu \B^{\mu\nu}}{-\Box}
\ri ] \del_\alpha 
 \lef [ A_\beta + \C_\beta + 
\fr{\del_\lambda \B^{\beta\lambda}}{-\Box}
\ri ] \ri ] \ri \}
\label{cff4}
\ee
The last term in the exponent in the above expression yields three new
contributions, which we call respctively $\overline{BB}$, $\overline{BC}$ 
and $\overline{BA}$.
The $\overline{BB}$ term is given by
\be
\overline{BB} = - \fr{ib^2 e^2}{16\pi} \int_{x_2,L}^{y_2} d\xi_\lambda
\int_{x_2,L}^{y_2} d\eta_\gamma 
\lef [ \epsilon^{\gamma\lambda\rho} \Box \del_\rho -
\epsilon^{\gamma\nu\rho}\del_\nu\del_\rho\del_\lambda \ri ]
\delta^3(\xi-\eta)\lef [\fr{1}{(-\Box)^2}  \ri]
\label{bbf}
\ee
The second term identically vanishes. The first term is also zero because
$d\xi^\lambda \parallel d\eta^\gamma$ for $\xi =\eta$. We therefore have
$\overline{BB} =0$.
Let us consider now $\overline{BC}$. Inserting (\ref{b}) and (\ref{c})
in (\ref{cff4}) we immediately find that $\overline{BC} = BD$ which is 
given by (\ref{bd}) and (\ref{bd1}). Finally the $\overline{BA}$ term.
Using (\ref{b}) and a trivial identity involving the Levi-Civita tensors,
we find
\be
\overline{BA} = -i\fr{e^2}{32\pi} \int d^3z \emab A_\mu (2 \B_{\alpha\beta})
\label{ba}
\ee
Inserting the results for these three terms in (\ref{cff4}), we find
$$
<\sigma \mu \sigma^\dagger \mu^\dagger> =
 Z^{-1} \int DA_\mu \expo 
\lef[ -\fr{i(\theta+\fr{e^2}{8\pi})}{4}
\emab A_\mu \lef (F_{\alpha\beta} + 2 \B_{\alpha\beta} +
2 \C_{\alpha\beta}\ri) 
\ri ]   \ri \} 
$$
\be
\times  \exp\{ \overline{BC} \}
\label{cff4}
\ee
Using the results of Section 2.3, we immediately find
$$
<\sigma(x_1) \mu_R(x_2) \sigma^\dagger(y_1)\mu_R^\dagger(y_2)> = \exp \lef\{ 
(\theta+\fr{e^2}{4\pi}) \fr{ab}{4\pi} \lef [\fr{1}{|y_1-x_2|} 
+ \fr{1}{|y_2 -x_1|}
\ri.\ri.
$$
\be
\lef.\lef.
-\fr{1}{|y_2-y_1|} - \fr{1}{|x_2 - x_1|} \ri ] 
 -i\fr{b^2 (\theta + \fr{e^2}{8\pi})}{4\pi} [ arg(\vec x_2-\vec y_2) 
 + arg(\vec y_2-\vec x_2) ]
\ri \}
\label{cff5}
\ee
This is our final result for the mixed correlation function in Chern-Simons
theory coupled to fermions in leading $\fr{1}{M}$ approximation.
		      
We can introduce a composite operator $\varphi$ bearing both charge and 
magnetic flux, through
\be
\varphi(x) = \lim_{x_1 \rightarrow x_2} \sigma(x_1) \mu_R(x_2)
\exp \lef\{ \fr{(\theta+\fr{e^2}{4\pi})}{4\pi |x_1 - x_2|}\ri \}
\label{fi}
\ee
The correlation functions of $\varphi$ can be obtained from (\ref{cff5}) and
(\ref{fi}), giving
\be
<\varphi(x) \varphi^\dagger(y)> = \exp \lef\{
(\theta+\fr{e^2}{4\pi}) \fr{ab}{2\pi} \lef [\fr{1}{|y-x|} 
 \ri ] 
 -i\fr{b^2 (\theta + \fr{e^2}{8\pi})}{4\pi} [ arg(\vec x-\vec y) 
 + arg(\vec y-\vec x)]
\ri \}
\label{cff6}
\ee
The multivaluedness of this correlation function, 
which is a consequence of the 
$arg$ functions is a reflection of the nontrivial commutation relation 
of the $\varphi$ field with itself \cite{vor,kc,ap}. Each sheet of
(\ref{cff6}) corresponds to a certain ordering of operators on the l.h.s..
The magnitude of the
jump from one sheet to another in (\ref{cff6}) indicates that $\varphi$
is an anyon operator with statistics 
$s = \fr{b^2 (\theta + \fr{e^2}{8\pi})}{4\pi}$.

\subsection{Physical States. Charge and Magnetic Flux Condensation}

Let us analize now the consequences of imposing the physical condition
(\ref{cons1}) on the charge and magnetic flux carrying states introduced
above. We immediately see that the $|\sigma>$ and $|\mu>$ states are both
unphysical. The anyon operator, however, does create physical 
states in the Chern-Simons theory with fermions, provided we choose the 
constants $a$, $b$ and $\theta$ in such a way that (\ref{cons1}) is
fulfilled. This happens when 
\be
e^2\  a\  {\cal M} = b
\label{rel}
\ee
For the special choice $a^{-1}= e \theta {\cal M}$ when $\sigma$
carries one unit of charge, we see that the condition for physical
states becomes $\theta b=e$.

The $\varphi$-states are usually anyonic. The only exception
occurs for the special value of $\theta = -\fr{e^2}{8\pi}$, 
for which the last term
in (\ref{cff6}) vanishes and $\varphi$ is bosonic
\be
<\varphi(x) \varphi^\dagger(y)> =
\exp \lef\{\fr{e^2ab}{16\pi |x-y|}\ri\}
\label{cff7}
\ee

It is quite interesting to study the long distance behavior of the 
$\varphi$ correlation function. From (\ref{cff7}) we see that in the 
bosonic case a unique limit exists, namely
\be
<\varphi(x) \varphi^\dagger(y)> 
{\stackrel{|\vec x-\vec y|\rightarrow \infty}{\longrightarrow}} 1
\label{longd}
\ee
This implies the condensation of charge and magnetic flux.

In the anyonic case the correlation function is multivalued, the
many branches of it corresponding to various posible orderings of the 
scalar operators on the left hand side \cite{vor,kc,ap}. In this case 
one should have the condensate of bosonic bound states of anyons.
We conclude that the only physical states associated with the 
$\sigma$ and $\mu$ operators
in Chern-Simons theory with
fermions form a condensate of charge and magnetic flux.

\section{Final Remarks}

The formalism for the description of quantum states bearing magnetic 
flux and charge developed in this work in the framework of 
Chern-Simons theory may have some  
interesting applications in several
models for condensed matter systems which are are based on this field theory.
The condensation of charge and magnetic flux in the physical states
suggests interesting phenomenological implications.
The formulation of correlation functions in terms of vacuum functionals 
in the presence of special external fields is practical and
allows the obtainment of concrete results with relative ease. We intend to
pursue in the near future the exploration of the possible phenomenological
consequences of the results contained in this work, in connection with 
condensed matter systems. Another potentially interesting application 
would be in the Chern-Simons theory coupled to a Higgs field.

\vfill\eject
				      
\appendix

\renewcommand{\thesection}{\Alph{section})}

\renewcommand{\theequation}{\Alph{section}.\arabic{equation}}

\section{ Appendix A}
\setcounter{equation}{0}

Let us demonstrate here Eq.(\ref{ds}) and thereby establish the path 
independence of (\ref{cf1}). Inserting (\ref{b}) in (\ref{cf1}) 
and integrating over $z$, we 
immediately see that the exponent in (\ref{cf1}) is in the form
\be
S_{CS}[\B_{\mu\nu}] =  S_{CS} - i b\theta \int_{x_i,L}^\infty d\xi^\mu A_\mu
\label{scs}
\ee
where $x_i$ is either $x$ or $y$.
Making the change of variable (\ref{ome}) in the above expression, we
get
\be
\Delta S_{CS}[\B_{\mu\nu}]= i \int d^3z \emab A_\mu \del_\alpha
\Omega_\beta
\label{ds1}
\ee
The $\Omega$ term coming from the last term in (\ref{scs}) vanishes
because it is proportional to 
$$
\int_{S(L,L')} d^2\xi_\mu \int_{x,L}^\infty d\eta_\mu \delta^3(\xi- \eta)=0
$$
This is zero because $d^2\xi^\mu \bot d\eta_\mu$ for $\xi=\eta$. Also the 
$\Omega-\Omega$ term coming from the first term in (\ref{scs}) vanishes
because it is proportional to
$$
\int_{S(L,L')} d^2\xi_\mu \int_{S(L,L')}d^2\eta_\beta \emab
\del_\alpha \delta^3(\xi- \eta)=0 
$$
This is zero because $d^2\xi_\mu \parallel d^2\eta_\beta$ for $\xi=\eta$.

Inserting $\Omega_\mu$ in (\ref{ds1}) and integrating over $z$, we get
$$
\Delta S_{CS}[\B_{\mu\nu}] = ib\theta \int_{S(L,L')} d^2\xi_\mu
\emab  \del_\alpha A_\beta
\label{ds2}
$$
\be
= -ib\theta \oint _{L'-L} d\xi^\mu A_\mu = S_{CS}[\B(L')] - S_{CS}[\B(L)]
\label{ds3}
\ee
This demonstrates (\ref{ds}).

\section{ Appendix B}
\setcounter{equation}{0}

Let us demonstrate here Eq.(\ref{ds5}) and thereby establish the path 
independence of (\ref{cf2}). Inserting (\ref{c}) in (\ref{cf2}) 
and integrating over $z$, we 
immediately see that the exponent in (\ref{cf2}) is in the form
\be
S_{CS}[\C_{\mu\nu}] =  S_{CS} + i a\theta \int_{x_i,L}^\infty d\xi^\mu
\emab \del_\alpha A_\beta
\label{scs1}
\ee
where $x_i$ is either $x$ or $y$.
Making the change of variable (\ref{omet}) in the above expression, we
get for $\Delta S_{CS}[\B_{\mu\nu}]$ an expression identical to (\ref{ds1}),
with $\tilde\Omega$ instead of $\Omega$. Again the 
$\tilde\Omega$-$\tilde\Omega$ and 
$\tilde\Omega$-$\C_{\mu\nu}$ terms vanish for the 
same reasons as in the previous
Appendix. Inserting (\ref{omet}) in the place of $\Omega_\mu$ in (\ref{ds1})
we get
$$
\Delta S_{CS}[\C_{\mu\nu}] = ia\theta \oint_{L'-L} d\xi_\mu
\emab  \del_\alpha A_\beta
\label{ds3}
$$
\be
S_{CS}[\C(L')] - S_{CS}[\C(L)]
\label{ds4}
\ee
This demonstrates (\ref{ds5}).

\section{ Appendix C}
\setcounter{equation}{0}

Let us demonstrate here Eq. (\ref{bb1}). We start by using the identity
$$
\emab \del_\alpha \Phi_\beta = \del^\mu \lef [ \fr{1}{|x|} \ri ]
$$
or
\be
\epsilon^{\mu\nu\alpha} \del_\alpha \lef [ \fr{1}{|x|} \ri ] =
\del^\mu \Phi^\nu - \del^\nu \Phi^\mu 
\label{id}
\ee
where
\be
\Phi^\mu = \fr{1 - \cos \theta}{r \sin \theta} \hat \varphi^\mu
\label{fi}
\ee
in spherical coordinates. Inserting (\ref{id}) in (\ref{bb}) we get
\be
BB =  \fr{ib^2\theta}{8\pi} \lef [ 
\int_{x_2,L}^{y_2} d\eta_\mu 
\lef [ \Phi^\mu (y_2-\eta) + \Phi^\mu (\eta-y_2) \ri ] 
- \int_{x_2,L}^{y_2} d\xi_\mu
\lef [ \Phi^\mu (x_2-\xi) + \Phi^\mu (\xi-x_2) \ri ] \ri ]
\label{bb2} 
\ee
Now, observing that 
\be
\Phi^\mu (x-\xi) + \Phi^\mu (\xi-x) = 2 \del^\mu arg(\vec x-\vec \xi)
\label{id1}
\ee
we immediately establish (\ref{bb1}) upon integration over $\xi$ and
$\eta$ in (\ref{bb2}).

\section{ Appendix D}
\setcounter{equation}{0}

Let us demonstrate here that the operator $\mu$, given by (\ref{muf1})
commutes with the charge operator. The first term in the exponent in
(\ref{muf1}) obviously commutes with 
the charge density $j^0$. The second term is
\be
J = \int d^3z int^\infty_{x,L} d\xi_\lambda \epsilon^{\lambda\mu\nu}
j_\mu(z) \del^{(\xi)}_\nu F(z-\xi)
\label{jo}
\ee
where
\be
F(x) = \fr{1}{-\Box} = \fr{1}{4\pi |x|}
\label{efe}
\ee
For convenience, we are working in euclidean space.
Inserting the identity (\ref{id}) in 
(\ref{jo}) and integrating in $\xi$ we get
\be
J = - \int d^3z j^\mu(z) \Phi_\mu (z-x)
\label{j1}
\ee
where $\Phi^\mu$ is given by (\ref{fi}). The piece corresponding to the last
term in (\ref{id}) vanishes because we can always choose $L$ such that 
$d\xi^\mu$ is orthogonal to $\Phi^\mu$. Since $j^\mu$ must be invariant
under reversal about the $z^3=0$ plane as a consequence of its invariance
under time-reversal in Minkowski space, 
only the first term of $\Phi_\mu$ contributes to
(\ref{j1}) and we can write
\be
J = - \int d^3z j^i(z)\del_i\phi (z-x) = - \int d^3z \del_0 j^0(z)\phi (z-x)
\label{j2}
\ee
where we integrated by parts and used the conservation of current.
It is now clear that $J$ also commutes with the charge density $j^0$ and
therefore so does the operator $\mu$.

\vfill\eject


\begin{thebibliography}{99}


\bibitem{nv} E.C.Marino, {\it Int. J. Mod. Phys.} {\bf A10} (1995) 4311

\bibitem{ap}
E.C.Marino, {\it Phys. Rev.} {\bf D38} (1988) 3194;  
{\it Ann. of Phys. (NY) } {\bf 224 } (1993) 225;

\bibitem{evora} E.C.Marino, {\it ``Dual Quantization of Solitons''} in
Proceedings of the NATO Advanced Study Institute, Applications of Statistical
and Field Theory Methods in Condensed Matter, (D.Baeriswyl, A.Bishop and
J.Carmelo, Eds.) Plenum, NY, 1992.

\bibitem{rede}J.Fr\" ohlich and P.A.Marchetti, {\it Lett. Math. Phys.}
{\bf 16} (1988) 347; {\it Commun. Math. Phys. } {\bf 121} (1989) 177;
J.Fr\" ohlich, F.Gabiani and P.A.Marchetti, in ``Proceedings, Banff
Summer School in Theoretical Physics'' (H.C.Lee, Ed.)

\bibitem{cont} G.W.Semenoff and P.Sodano, {\it Nucl. Phys.} {\bf B328}
(1989) 753; M.L\"uscher, {\it Nucl. Phys.} {\bf B326} (1989) 557;
R.Jackiw and S.Y.Pi, {\it Phys. Rev.} {\bf D42} (1990) 3500;
A.Kovner, B.Rosenstein and D.Eliezer, {\it Nucl. Phys.} {\bf B350}
(1991) 325; {\it Mod. Phys. Lett.} {\bf A5} (1990) 2733;
V.F.M\" uller, {\it Z.Phys.} {\bf C51} (1991) 665


\bibitem{oop} E.C.Marino, {\it ``Local Charged Operators in 2+1 D 
Maxwell-Type Theories''}, in preparation

\bibitem{dr} S.Deser and A.N.Redlich, {\it Phys. Rev. Lett.} {\bf 61} (1988)
 1541

\bibitem{vor} E.C.Marino and J.A.Swieca, {\it Nucl. Phys.} {\bf B170}
(1980) 175

\bibitem{kc} L.P.Kadanoff and H.Ceva, {\it Phys. Rev.} {\bf B3} (1971) 3918





 









\end{thebibliography}
\end{document}